\documentclass[sn-standardnature]{sn-jnl}

\usepackage{subfig} 
\usepackage{doi}
\usepackage{pdflscape}	
\usepackage[none]{hyphenat}

\newcommand{\araa}{Annu. Rev. Astron. Astrophys.}   
\newcommand{\ac}{Astron. Comput.} 
\newcommand{\apj}{Astrophys. J.}   
\newcommand{\apjl}{Astrophys. J. Lett.}   
\newcommand{\apjs}{Astrophys. J. Suppl. Ser.}   
\newcommand{\aap}{Astron. Astrophys.}   
\newcommand{\jai}{J. Astron. Instrum.}  
\newcommand{\mnras}{Mon. Not. R. Astron. Soc.}   
\newcommand{\nat}{Nature} 
\newcommand{\nastro}{Nat. Astron.} 
\newcommand{\prd}{Phys. Rev. D}   
\newcommand{\pasa}{Publ. Astron. Soc. Aust.}   
\newcommand{\raa}{Res. Astron. Astrophys.} 
\newcommand{\ssr}{Space Sci. Rev.}   
\newcommand{\sciadv}{Sci. Adv.} 
\newcommand{\advsr}{Adv. Space Res.} 
\newcommand{\ijmphys}{Int. J. Mod. Phys. D} 

\newcommand{\ulpolong}{ASKAP\,J193505.1\ensuremath{+}214841.0}
\newcommand{\ulpo}{ASKAP\,J1935\ensuremath{+}2148}
\newcommand{\sgr}{SGR\,1935\ensuremath{+}2154}
\newcommand{\gpm}{GPM\,J1839\ensuremath{-}10}
\newcommand{\glx}{GLEAM-X\,J162759.5\ensuremath{-}523504.3}
\newcommand{\gcrt}{GCRT\,J1745\ensuremath{-}3009}
\newcommand{\psr}{PSR\,J1107\ensuremath{-}5907}

\newcommand{\dmunits}{\ensuremath{{\rm pc \, cm^{-3}}}}

\newcommand{\rmunits}{\ensuremath{{\rm rad \, m^{-2}}}}

\def\arcsec{\hbox{$^{\prime\prime}$}}
\newcommand\farcs{\hbox{$.\!\!^{\prime\prime}$}}
\newcommand\arcmin{\hbox{$.\!\!^{\prime}$}}
\def\Msun{\ifmmode{{\rm M}_\odot}\else${\rm M}_\odot$\fi}
\begin{document}

\title[An emission state switching 54-minute period radio transient]{An emission state switching radio transient with a 54 minute period}


\author*[1,2]{\fnm{M.} \sur{Caleb}}\email{manisha.caleb@sydney.edu.au}
\equalcont{These authors contributed equally to this work.}

\author*[3]{\fnm{E.} \sur{Lenc}}\email{Emil.Lenc@csiro.edu.au}
\equalcont{These authors contributed equally to this work.}

\author[4]{\fnm{D. L.} \sur{Kaplan}}
\author[1,2]{\fnm{T.} \sur{Murphy}}
\author[5]{\fnm{Y. P.} \sur{Men}}
\author[6]{\fnm{R. M.} \sur{Shannon}}
\author[7]{\fnm{L.} \sur{Ferrario}}
\author[8,9]{\fnm{K. M.} \sur{Rajwade}}
\author[10]{\fnm{T. E.} \sur{Clarke}}
\author[10]{\fnm{S.} \sur{Giacintucci}}
\author[11]{\fnm{N.} \sur{Hurley-Walker}}
\author[12]{\fnm{S. D.} \sur{Hyman}}
\author[3]{\fnm{M. E.} \sur{Lower}}
\author[11]{\fnm{Sam} \sur{McSweeney}}
\author[13,14]{\fnm{V.} \sur{Ravi}}
\author[5]{\fnm{E. D.} \sur{Barr}}
\author[15]{\fnm{S.} \sur{Buchner}}
\author[6]{\fnm{C. M. L.} \sur{Flynn}}
\author[8]{\fnm{J. W. T.} \sur{Hessels}}
\author[6]{\fnm{M.} \sur{Kramer}}
\author[1]{\fnm{J.} \sur{Pritchard}}
\author[16]{\fnm{B. W.} \sur{Stappers}}

\affil[1]{\orgdiv{Sydney Institute for Astronomy, School of Physics}, \orgname{The University of Sydney}, \orgaddress{\city{Sydney}, \postcode{2006}, \state{NSW}, \country{Australia}}}

\affil[2]{\orgname{ARC Centre of Excellence for Gravitational Wave Discovery (OzGrav)}, \orgaddress{ \city{Hawthorn}, \postcode{3122}, \state{Victoria}, \country{Australia}}}

\affil[3]{\orgname{Australia Telescope National Facility, CSIRO, Space \& Astronomy}, \orgaddress{\street{PO Box 76}, \city{Epping}, \postcode{1710}, \state{NSW}, \country{Australia}}}

\affil[4]{\orgname{Center for Gravitation, Cosmology, and Astrophysics, Department of Physics, University of Wisconsin-Milwaukee}, \orgaddress{\street{P.O. Box 413}, \city{Milwaukee}, \postcode{53201}, \state{WI}, \country{USA}}}

\affil[5]{\orgname{Max-Planck-Institut f{\"u}r Radioastronomie}, \orgaddress{\street{D-53121}, \city{Bonn}, \postcode{53121}, \country{Germany}}}

\affil[6]{\orgname{Centre for Astrophysics and Supercomputing, Swinburne University of Technology}, \orgaddress{\street{John Street}, \city{Hawthorn}, \postcode{3122}, \country{Australia}}}

\affil[7]{\orgname{Mathematical Sciences Institute, The Australian National University}, \orgaddress{ \city{Canberra}, \state{ACT}, \postcode{2601}, \country{Australia}}}

\affil[8]{\orgname{ASTRON, the Netherlands Institute for Radio Astronomy, Oude Hoogeveensedĳk 4}, \orgaddress{\city{PD Dwingeloo}, \postcode{7991}, \country{The Netherlands}}}

\affil[9]{\orgname{Astrophysics, University of Oxford, Denys Wilkinson Building}, \orgaddress{Keble Road, Oxford}, \postcode{OX1 3RH}, \country{UK}}

\affil[10]{\orgname{Remote Sensing Division, U.S.\ Naval Research Laboratory}, \orgaddress{\city{Washington D.C.}, \postcode{20375}, \country{USA}}}

\affil[11]{\orgname{International Centre for Radio Astronomy Research, Curtin University}, \orgaddress{\street{Kent street}, \city{Bentley WA}, \postcode{6102}, \country{Australia}}}

\affil[12]{\orgname{Department of Engineering and Physics, Sweet Briar College}, \orgaddress{\city{VA}, \postcode{24595}, \country{USA}}}

\affil[13]{\orgname{Cahill Center for Astronomy and Astrophysics, MC 249-17 California Institute of Technology}, \orgaddress{\city{Pasadena}, \postcode{91125}, \country{USA}}}

\affil[14]{\orgname{Owens Valley Radio Observatory, California Institute of Technology, Big Pine}, \orgaddress{\city{Big Pine}, \postcode{CA 93513}, \country{USA}}}

\affil[15]{\orgname{South African Radio Astronomy Observatory (SARAO), 2 Fir Street, Observatory}, \orgaddress{\city{Cape Town}, \postcode{7925}, \country{South Africa}}}

\affil[16]{\orgdiv{Jodrell Bank Centre for Astrophysics, Department of Physics and Astronomy}, \orgname{The University of Manchester}, \orgaddress{\street{Oxford road}, \city{Manchester}, \postcode{M13 9PL}, \country{United Kingdom}}}



\abstract{Long-period radio transients are an emerging class of extreme astrophysical events of which only three are known. These objects emit highly polarised, coherent pulses of typically a few tens of seconds duration and minutes to $\sim$hour-long periods. While magnetic white dwarfs and magnetars, either isolated or in binary systems, have been invoked to explain these objects, a consensus has not emerged. Here we report on the discovery of \ulpolong{} (henceforth \ulpo{}) with a period of 53.8~minutes exhibiting three distinct emission states – a bright pulse state with highly linearly polarised pulses with widths of 10--50 seconds; a weak pulse state which is about 26 times fainter than the bright state with highly circularly polarised pulses of widths of approximately 370 milliseconds; and a quiescent or quenched state with no pulses. The first two states have been observed to progressively evolve over the course of 8~months with the quenched state interspersed between them suggesting physical changes in the region producing the emission. A constraint on the radius of the source for the observed period rules out a magnetic white dwarf origin. Unlike other long-period sources, \ulpo{} is the first to exhibit drastic variations in emission modes reminiscent of neutron stars. However, its radio properties challenge our current understanding of neutron star emission and evolution.}


\maketitle

\ulpo{} was serendipitously discovered during a target of opportunity observation of the gamma-ray burst GRB~221009A with the Australian Square Kilometre Array Pathfinder (ASKAP) telescope. Bright pulses of radio emission from \ulpo{} were seen on 2022-10-15 using a fast-imaging technique on images with 10~s integration times (Methods). \ulpo{} is located at RA (J2000) = $19^h35^m05.126^s \pm 1.5''$ and Dec (J2000) = $+21^d48'41.047'' \pm 1.5''$, which is coincidentally 5\arcmin6 from the magnetar \sgr{} and sits on the edge of the supernova remnant in which \sgr{} is centred. 
The observation lasted $\sim6$\,h revealing 4 bright pulses lasting 10 -- 50 seconds in the images, with the brightest peak pulse flux density measuring 119\,mJy. Inspection of the light curves of the pulses revealed a tentative period of $\sim 54$~minutes.

In addition to a weak detection in an archival ASKAP observation, the source was consistently detected in four follow-up observations.
A summary of all observations with ASKAP is presented in Extended Data Table 1. Overall, the pulses are visible across the whole bandpass corrected observing band of 288~MHz leading to a spectral index estimation of $\alpha \approx +0.4\pm0.3$ at 887.5~MHz. However, a dispersion measure (DM) constraint/estimate was not possible due to the coarse time resolution of 10~s. We quantify the pulses to be $>90\%$ linearly polarised --- implying strongly ordered magnetic fields, with a rotation measure (RM) of $+159.3\pm0.3$\,\rmunits\, calculated using the RM synthesis method \citep{bdb2005}. In comparison, the RM and DM of \sgr{} are $\sim+107$\,\rmunits and $\sim330$\,\dmunits, respectively \citep{zxz+23}. The RM of \ulpo{} is consistent with the contribution from the smoothed Galactic foreground \citep{ojr+12} and with those of nearby pulsars (\url{https://www.atnf.csiro.au/research/pulsar/psrcat/}), precluding the presence of a significant RM imparted at the source.

Following the discovery, we conducted simultaneous beamformed and imaging follow-up observations at 1284~MHz with the MeerKAT radio interferometer (Methods). Two pulses were detected in both the beamformed and imaging data in two independent observations (Extended Data Table 1). 
The initial estimate of period allowed us to predict the times of arrival of future pulses at the same rotational phase of \ulpo{}, and the MeerKAT pulses are observed to arrive within 319~ms of the predicted times (i.e. within $10^{-4}$ of a period). The times of arrival of all the ASKAP and MeerKAT detections were used to determine a phase-connected timing solution with a period $P$ of $3225.313\pm0.002$~seconds (Methods and Figure \ref{fig:lightcurves}), and an upper limit on the period derivative, $\dot{P}$ of $\lesssim (1.2\pm1.5) \times 10^{-10}$~s\,$\rm{s}^{-1}$ with a $1\sigma$ error. The location of \ulpo{} in the $P-\dot{P}$ parameter space, which is commonly used to classify different sorts of pulsars, is consistent with other known long-period sources (Extended Data Figure 1). \ulpo{} is seen to reside in the pulsar death valley where detectable radio signals are not expected, challenging currently accepted theories of radio emission via spin-down (Extended Data Figure 1). The radio properties of \ulpo{} are presented in Table \ref{tab:params}.

A first single pulse was detected by the MeerTRAP realtime detection system (Methods) on 2023-02-03 with a DM of $145.8\pm3.5$\,\dmunits\, and a width of $\sim370$\,ms (Extended Data Table 1), which is $\sim135\times$ narrower than the brightest ASKAP pulse. Contrary to the ASKAP duty cycle of 1.5\%, the narrow width of the MeerKAT pulse results in a duty cycle of only 0.01\%. The average DM inferred distance based on the NE2001 \citep{ne2001} and YMW16 \citep{ymw16} Galactic electron density models places \ulpo{} at a distance of 4.85~kpc (Table \ref{tab:params}). The detection is accompanied by weak pre- and post-cursor pulses as seen in Figure \ref{fig:pulse-profiles}. The data recorded to disk with the PTUSE backend (Methods) did not reveal a broader underlying emission envelope similar to the wide pulse widths seen in the ASKAP detections. The corresponding MeerKAT 2-s resolution image (the shortest possible timescale) revealed a single 9~mJy detection, which is $\sim26\times$ fainter than the brightest ASKAP pulse. The pulse was localised to RA (J2000) = $19^h35^m05.175^s \pm 0.3''$ and Dec (J2000) = $+21^d48'41.504'' \pm 0.6''$ which is consistent with the ASKAP coordinates. Throughout the paper, the flux densities quoted for the MeerKAT data are from the images as the beamformed data are only polarisation- and not flux-calibrated. Unlike the ASKAP pulses, our MeerKAT detection revealed a substantial circular polarization fraction exceeding 70\%, coupled with a linear polarization fraction of $\sim40\%$. We did not find evidence for Faraday conversion (Methods). In addition to being spatially coincident, the measured RM of $+159.8 \pm 0.3$ \rmunits\, agrees with that measured for the ASKAP detections giving us added confidence that despite the drastically different pulse widths, the ASKAP and MeerKAT detections were produced by the same object. 

The second MeerKAT detection was made on 2023-05-08 with a flux density of 2.9~mJy averaged over 2~s (see Extended Data Table 1). This burst was also $\sim370$~ms wide and highly circularly polarised with no broader emission envelope, but the lack of sufficient signal-to-noise precluded a reliable RM estimation. Both MeerKAT pulses are visible across the whole 856~MHz band with a spectral index estimate of $\alpha \approx -1.2\pm0.1$ at 1284~MHz. Such variations in spectral indices have been observed in both pulsars and radio magnetars exhibiting different emission states \citep{ljs+21, bmm22}.
Combining all the ASKAP and MeerKAT observations (see Extended Data Table 1), we see that the source is not detectable in every single observation indicating intermittency, potential nulling where the pulsed emission temporarily ceases or becomes undetectable, or drastic variations in flux density. Using epochs 1 to 17 in Extended Data Table 1, we estimate the source to be in a quenched or quiescent state $\sim 40-50\%$ of the time, at MeerKAT and ASKAP.

Observationally, \ulpo{} appears to exhibit three emission states: 
\begin{enumerate}
    \item the strong pulse mode consisting of 15 bright, tens of seconds wide, and highly linearly polarised pulses as seen with ASKAP; 
    \item the weak pulse mode characterized by 2 faint, hundreds of milliseconds wide, and highly circularly polarised pulses as seen with MeerKAT;
    \item the completely nulling or quiescent mode as seen with both telescopes.
\end{enumerate}

\noindent We consider two possible scenarios for the observed differences in properties of the ASKAP and MeerKAT bursts. \\

\underline{\textit{Scenario 1}}: The ASKAP pulses could span only a small fraction (i.e. a few hundreds of milliseconds) of the shortest possible time resolution of 10~s. This scenario would imply that the source only produces sub-second duration pulses. In Figure \ref{fig:lightcurves} we see the flux densities of detections from 10~s ASKAP images to gradually rise and fall resulting in an almost Gaussian-like pulse profile. This distribution of the flux densities makes it unlikely for the burst to be comprised of several consecutive millisecond duration pulses. However, it remains possible for sub-second timescale structure to be superimposed on the broader emission envelope.
    
\underline{\textit{Scenario 2}}: It is likely that there are different emission modes at play. The source was undetectable with ASKAP in all follow-up observations post 2022-11-05, until the first MeerKAT observation on 2023-02-03 with its 5 times better sensitivity. The MeerKAT pulses which would have been undetectable at ASKAP are analogous to the `quiet' pulse mode in PSR~B0823$+$26, and the `dwarf pulse' mode in PSR~B2111$+$46 \citep{cyh+23}. The pulses in these modes are generally undetectable in lower-sensitivity and/or low time-resolution observations such as with ASKAP. 
These weak pulses potentially exist between the nulling or quenched states of \ulpo{}. The location of the source at the edge of the supernova remnant in Extended Data Figure 2 makes it is difficult to determine exactly what background emission to subtract. Therefore we are unable to confirm the presence of persistent continuum radio emission which might be indicative of a wind nebula in either the MeerKAT or ASKAP data (Methods).

Collectively, the pulse widths, spectra and polarisation properties of the ASKAP and MeerKAT detections suggest different physical coherent processes even though they occur at roughly the same rotational phase. Coherent radio emission from rotating neutron stars is efficiently generated by the creation of electron-positron pairs in the magnetosphere. The rotational spin-down creates an electric potential at the polar cap, causing  pair production. Such charged plasma can emit radio waves which can be attributed to curvature radiation and inverse Compton scattering, and diverse magnetic field configurations in emission models, including dipolar, multipolar, and twisted fields along with vacuum gaps and space-charge-limited flows~\citep{rs75, zgd07}. Magnetically powered neutron stars on the other hand generate coherent radio emission through decaying magnetic fields \citep{kb17}. Extended Data Figure 3 shows the manifestation of the physics underlying coherent and incoherent emitters, and indicates a coherent emission mechanism (brightness temperatures between $10^{14}$~K and $10^{16}$~K) being responsible for both the ASKAP and MeerKAT detections of \ulpo{}.

When comparing with other known long-period sources, \ulpo{} appears to be similar to \glx{} \citep{hwzb+22} and \gpm{} \citep{hwrm+23} albeit with a period that is 4 times longer but with a duty cycle not too dissimilar. \glx{} was active for only three months, while \gpm{} has remained active for over three decades \citep{hwzb+22, hwrm+23}. Despite searches across radio data from the Giant Metrewave Radio Telescope (GMRT), Very Large Array (VLA) and VLA Low-band Ionosphere and Transient Experiment (VLITE) spanning 2013 to 2023, no significant pulses from \ulpo{} were detected. We note that \ulpo{} shares similarities with the Galactic center radio transient (GCRT), or ``Burper", \gcrt{}. At the time of its discovery, \gcrt{} exhibited $10$-minute wide pulses with a periodicity of 77~minutes \citep{hlj+05}, but subsequent observations revealed narrower and weaker pulses spanning two minutes \citep{hrp+07}. Varying circular polarisation was also found in one of the pulses \citep{rhp+2010}. The similarities in the periods and the different emission states imply that, \ulpo{} could be a bridge between \gcrt{}, \glx{} and \gpm{}. 

Due to the proximity of \sgr{}, there are numerous archival high-sensitivity X-ray observations at the position of \ulpo{} (see Methods and Extended Data Table 2). We focused on observations with the \textit{Chandra X-ray Observatory} and \textit{Neil Gehrels Swift Observatory} \citep{swift}. Using a combination of the more sensitive \textit{Chandra} observations, we did not detect any X-ray source at the location of \ulpo{}. This corresponds to a luminosity limit of $\approx 4\times 10^{30}d_{4.85}^2\,{\rm erg\,s}^{-1}$ (Methods) for a blackbody spectral model and a powerlaw spectral model for non-thermal emission from a neutron star. This is below the X-ray luminosities of most but not all rotation-powered pulsars and magnetars~\citep{2022A&A...658A..95V, 2014ApJS..212....6O} and is comparable to the X-ray luminosities of other long period radio transients ($\leq 10^{32-33}\,{\rm erg\,s}^{-1}$) \citep{hwzb+22, hwrm+23}.
We also searched for flaring activity in \ulpo{} using \textit{Swift}, with  291 individual visits using the X-ray Telescope (XRT; \cite{xrt}) lasting 5--2600\,s from 2010 December through 2022 December for a total exposure time of 302.4\,ks (exposure-corrected). We see no sources in the summed dataset at the position of \ulpo{} (Methods). The repeated visits with \textit{Swift} allow us to rule out any flaring behavior during this period. 

Archival 300\,s exposures in the $J$, $H$, and $K_s$ bands (1.2\,$\mu$m, $1.6\,\mu$m, and $2.1\,\mu$m) with the Very Large Telescope (VLT) using the near-infrared HAWKI \citep{hawki} imager showed a source within the conservative $1.5\arcsec$ ASKAP error radius of \ulpo{} (Figure \ref{fig:vlt_archival}). This source with $J=18.4\pm0.1$\,mag, $H=17.3\pm0.1\,$mag, and $K_s=17.1\pm0.1$\,mag (Vega) is cataloged as PSO~J293.7711+21.8119 in Data Release 2 of the Pan-STARRS1 (PS1) $3\pi$ survey \citep{ps1}. We compute the chance of finding a source randomly (drawn from the background) in the $K_s$ image, which has the highest source density, with magnitude brighter than or equal to this value to be 5\% (i.e. $\sim2\sigma$ association) given the crowded nature of the field. However, to confidently rule out the association we obtained a spectrum of PSO~J293.7711$+$21.8119 in the 3,200--10,000 Angstroms wavelength range with the Low Resolution Imaging Spectrometer (LRIS) at the Keck telescope in Hawaii (see Extended Data Figure 4). The calibrated spectrum is a red continuum devoid of discernible emission or absorption lines. The combination of the VLT magnitudes and the spectral characteristics suggests it is a L/T-dwarf star. Gaia DR1 parallaxes for known L and T dwarfs give $J$-band absolute magnitudes $M_J$ in the range $10 < M_J < 16$ \citep{smc+17}, which implies a distance of less than 0.5~kpc for an apparent magnitude of $J=18.4$. Because at 4.85~kpc, such a star would be undetectable, we conclude that PSO~J293.7711$+$21.8119 is a foreground star that is unlikely to be associated with \ulpo{}.

The observed period and emission of \ulpo{} could be explained by a rotating magnetic white dwarf (MWD) emitting coherent radio emission like a neutron-star pulsar \citep{rhp+23}. 
MWDs can either be isolated or in interacting binaries. There are $\sim600$ known isolated MWDs with surface dipole magnetic fields up to $10^9$~G and $\sim 200$ in interacting binaries with magnetic fields up to a few $10^8$~G \citep{fwk+20}. 
Although radio emission from isolated MWDs has never been detected, despite searches for possible counterparts in large area radio surveys \citep[e.g.,][]{Barrett2020, pcs+24}, we now entertain this possibility and derive the parameters that would be required to explain the radio emission of \ulpo{}. If \ulpo{} is an isolated rotation-powered MWD, the measured $P$ and upper limit on $\dot{P}$ would yield a surface magnetic field strength and spin-down luminosity of a few $10^{10}$\,G and a few  $10^{31}$\,erg\,s$^{-1}$, respectively, for a dipolar magnetic field configuration, a magnetic inclination angle of $90^{\circ}$, and a moment of inertia of $10^{50}$~g~cm$^{2}$. Even though the currently known isolated MWDs have magnetic fields below $10^9$~G, it is theoretically possible for MWDs to have surface fields of up to a few $10^{13}$~G \citep{Franzon2015, Otoniel2019}. In this case, \ulpo{} would be the first MWD discovered to possess such a high magnetic field. 

The radius of the source can be related to rotational period and magnetic field strength to estimate the minimum radius of the source \citep{bwh+23} (Methods and Extended Data Figure 5). Even under the most conservative assumptions, we can rule out an isolated MWD origin if we presume that the magnetic field cannot exceed $10^9$~G, which is the maximum ever measured in a MWD. Similar considerations can also be applied to \glx{} and \gpm{}, and we conclude that it is highly unlikely that the radio emission from these sources can be interpreted in terms of an isolated rotation-powered MWD. However, coherent and highly polarised radio emission has been detected in cataclysmic variables (CVs) \citep{Barrett2020, Ridder2023} which are close binary systems containing a WD primary accreting matter from a low-mass M-dwarf companion. In all detection cases, the radio emission appears to arise from the lower corona of the magnetically active M-dwarf and is attributed to the electron cyclotron maser instability. The problem here is that the radio luminosities of CVs, in the range $10^{21}-10^{25}$ erg\,s$^{-1}$ \citep{Barrett2020}, would be too low to explain the emission of \ulpo{}. Hence, it is also highly unlikely that a CV could be responsible for the radio emission of \ulpo{}.

Assuming a neutron star origin, the period and upper limit on the period derivative correspond to a surface magnetic field strength and spin-down luminosity of a few $10^{16}$ G and a few $10^{26}$\,erg\,s$^{-1}$, respectively, for a dipolar magnetic field configuration, a magnetic inclination angle of $90^{\circ}$, and a moment of inertia of $10^{45}$~g~cm$^{2}$. It is unclear why a neutron star magnetar would still possess such a large magnetic field at this stage of its evolution, but explanations have been provided either in terms of the magnetic field's structure \citep[e.g.,][]{Pons2007, FW2008} or as due to a fall-back accretion disc \citep[e.g.,][]{Rea2022}. Similar to \glx{} \citep{hwzb+22} and \gpm{} \citep{hwrm+23}, the observed radio luminosity of \ulpo{} is much larger than the inferred spin-down luminosity, suggesting that alternative emission mechanisms must be involved to explain the radio emission of these long-period radio transients.

Achieving the observed duty cycle of approximately $\sim1$ percent in \ulpo{} necessitates a high degree of beaming, implying the generation and acceleration of relativistic particles -- a phenomenon which is generally more easily accommodated in neutron stars than in WDs. Remarkably, the isolated intermittent pulsar \psr{}, with $P \sim253$~ms, displays the same three distinct emission states as \ulpo{}\, \citep{yws+14}. The emission in both \psr{} and \ulpo{} alternates between distinct modes, each characterized by unique pulse profiles, polarization properties, and at times, varying intensities. 
The intricate interplay of magnetic fields, plasma flows, and the magnetospheric environment leads to the emergence of these different modes \citep{bms+82}. Instabilities within the pulsar magnetosphere may trigger transitions between these modes, contributing to the observed switching phenomenon. Additionally, changes in the geometry of the magnetic field configuration and the location of emission sites within the magnetosphere could influence the emitted radiation characteristics. Notably, \psr{} is close to the pulsar death line(s) (Extended Data Figure 1) \citep{CR93, ZHM2000}, beyond which radio emission is expected to cease. All these similarities seem to suggest that a neutron star-like emission mechanism is at play for \ulpo{}. 

In summary, we report the discovery of the long-period source \ulpo{}, which is unique compared to other known long-period sources by manifesting three distinctive emission states reminiscent of mode-switching pulsars. The strong pulse mode displays bright and linearly polarised pulses lasting tens of seconds, the weak pulse mode features faint and circularly polarised pulses lasting hundreds of milliseconds, and the completely nulling or quiescent mode exhibits an absence of pulses. These diverse emission states offer valuable insights into the magnetospheric processes and emission mechanisms at play within this object, with similarities to the radio pulsars \psr{}, PSR~B0823$+$26 and PSR~B2111$+$46. Given \glx{}'s brief 3-month activity, ongoing monitoring may unveil emission modes, similar to the those observed in \ulpo{}. We see that radio emission via pair production within dipolar magnetospheres present significant challenges \citep{rhp+23}. However, a large magnetic field can power the observed radio emission via the dissipation of energy due to magnetic re-connection events, higher order magnetic fields and untwisting of field lines due to plastic motion of the crust \citep{jones2003, lyutikov2015}. It would be prudent to study further whether such processes can persist for long timescales consistent with the long-term emission seen in a few long-period sources. Population-synthesis simulations incorporating various parameters such as masses, radii, beaming fractions, and magnetic field show that only a limited number of long period radio emitters of neutron star origins as expected to exist in the Galaxy \citep{rhp+23}. Conversely, in the WD scenario, a sizable population of long period emitters can be accounted for. Nonetheless, explaining the production of coherent radio emission remains a formidable task in either scenario \citep{rhp+23, tong23}.
While MWDs have been considered to be responsible for the radio emission observed in sources like \glx{} and \gpm{}, we have ruled out this possibility for \ulpo{}. Thus, it is much more likely for \ulpo{} to be an ultra-long period magnetar or neutron star either isolated or in a binary system. Continued monitoring of this source should  allow us to determine whether additional periods are present and the possible existence of a companion star. 

\clearpage



\clearpage

\section*{Methods}\label{sec6}

\section*{ASKAP}
\label{sec:askap}

The ASKAP array comprises 36 antennas, each equipped with a prime-focus phased array feed (PAF). Each PAF has 188 linearly-polarised receiving elements sensitive to frequencies between 0.7 and 1.8 GHz. The signal from each element is channelised to 1 MHz frequency resolution over a usable bandwidth of 288 MHz. The standard ASKAP hardware correlator produces visibilities on a 10~s timescale. During the detection of \ulpo{}, ASKAP was operated in the \texttt{square\_6x6} configuration with 1.05 deg pitch and 887.5 MHz central frequency. The pointing centre was chosen such that ASKAP's large field-of-view ($\sim30$ deg$^{2}$) would also encompass the magnetar \sgr\, known to have produced a burst with FRB like energies in 2020 \citep{brb+20, chimesgr1935} and several less energetic bursts since. 

The source was found during testing of a fast pulse-detection pipeline. The pipeline subtracts visibilities from neighbouring 10-second integrations and then images the result. The process effectively subtracts out quiescent emission from the field and only retains sources that change dramatically over a single integration. Since such sources are rare, most images are effectively dominated by thermal noise and so do not require computationally expensive deconvolution. Once the image is checked for signficant peaks above the noise, it is discarded to minimise storage requirements. The process is repeated for all time integrations and all 36 ASKAP beams. The processing initially found no detections in the first epoch of the ASKAP observation but found a significant pulse in beams 21, 22 and 23 of the second ASKAP epoch.

While the pulse-detection pipeline is reasonably effective for finding bright pulses it forgoes some sensitivity to weak pulses to minimise resource usage. Once we had discovered \ulpo{}, more traditional techniques were used to investigate the pulse. A deep model image was derived for the beam and subtracted from the visibility data. The data was then phase-shifted to the location of the source and dynamic spectra extracted (averaging over all baselines $>200.0$~m). This allowed weaker pulses to be detected and also allowed the linear polarisation properties to be analysed. This approach was also repeated for MeerKAT observations.

\section*{Murriyang}

The shortest ASKAP imaging observation (SBID 44918 in Extended Data Table 1), took place at the same time the 64\,m Parkes (Murriyang) radio telescope was pointed at the known radio-burst emitting magnetar \sgr. This observation was taken using the Ultra Wideband Low (UWL) receiver system, spanning a bandwidth of 704--4032\,MHz. The position of \ulpo{} was well within the 30 arcmin wide low-frequency part of the UWL beam. Only one weak pulse ($\sim9$ sigma or 10.7 mJy/beam) was detected from \ulpo{} in the ASKAP data but no detection was made in the Parkes data above a S/N of 8.

\section*{MeerKAT}
\label{sec:methods-mkt}

In the observations presented in this work, MeerKAT operated at L-band (0.86–1.71~GHz) in the c856M4k configuration where the correlations were integrated for 2-seconds before saving to disk. We used PKS J1939$-$6342 as the primary flux calibrator and bandpass calibrator, and PKS J2011$-$0644 as the phase calibrator. MeerKAT was also used to simultaneously perform beamformed observations in the incoherent and coherent modes using MeerTRAP. The MeerTRAP backend is the combination of two systems: the Filterbank and Beamforming User Supplied Equipment (FBFUSE), a multiple-beam beamformer \citep{barr2018, CBK+21}, and the Transient User Supplied Equipment (TUSE), a real-time transient detection instrument \citep{CSA+20}. 
FBFUSE applies the geometric and phase delays (obtained by observing a bright calibrator) before combining the data streams from the dishes into one incoherent beam and up to 780 coherent beams recording in total intensity, only. The beams can be placed at any desired location within the primary beam of the array, but are by default tessellated into a circular tiling centered on the boresight position, and spaced so that the response patterns of neighbouring beams intersect at the 25\% peak power point. The beams are then sent over the network to TUSE for processing.

The TUSE single pulse search pipeline searches for total intensity pulses in realtime at a sampling time of $306.24~\mu$s, up to a maximum boxcar width of 140~ms in the dispersion measure range of 0--5118.4 \dmunits\, at L-band. Only extracted candidate files are saved to disk for further investigation. Further details on the MeerTRAP backend can be found in \cite{rsw+20} and \cite{chr+22}. The Accelerated Pulsar Search User Supply Equipment (APSUSE) backend instrument was used to record total intensity data with 4096 frequency channels over a 856-MHz band centered at 1.284 GHz to disk for all beams at a sampling time of $76.56~\mu$s, for the entire duration of the observations. Additionally, full polarization data for the on-source beam were recorded to disk, but not searched in realtime, using the Pulsar Timing User Supplied Equipment (PTUSE) backend of the MeerKAT pulsar timing project MeerTime described in \cite{meertime2020}. These data were recorded in the PTUSE search mode with a sampling time of $38.28~\mu$s in the \texttt{psrfits} format.

To address the impact of baseline variation in the recorded data, we utilized the APSUSE off-source beam as a reference to eliminate the baseline of the on-source beam. Following this process, we conducted a search for single pulses using \texttt{TransientX} ({\url{https://github.com/ypmen/TransientX}) \citep{Men2024arXiv}, within a DM range of 120-160\,{\dmunits} and a maximum pulse width of one second. This search yielded the detection of the two pulses in the PTUSE data. Subsequently, we extracted the polarization profiles from the PTUSE data of these two pulses, having removed the baseline based on the off-source APSUSE beam.

\section*{Period estimation}

We generated times of arrival (ToAs) for each of the detections made with the ASKAP and MeeKAT observation in Extended Data Table 1. Manipulation of the data used the tools available in the \textsc{psrchive} package \citep{Hotan+2004}. The ToAs for the ASKAP detections are chosen to be the midpoints of the 10~s integrations they were detected in, with the error on the ToA equal to the duration of an non-detection in 10~s integrations immediately preceding, and succeeding the first and last detections of a pulse, respectively. The MeerKAT ToAs were obtained by using \textsc{psrchive}'s \textsc{pat} on the beamformed data to estimate the time of the peak flux. Given the variability in the morphology of individual pulses, and presumable jitter in the emission  measuring the pulse arrival times, we estimate the uncertainty on the arrival times to be the half width at half maximum of the widths of the pulses.
The initial timing analysis for both telescopes used the best known period and DM at the time and the position determined from the imaging.

Timing was done using \textsc{tempo2} \citep{tempo2} with the JPL DE436 planetary ephemeris (\url{https://naif.jpl.nasa.gov/pub/naif/JUNO/kernels/spk/de436s.bsp.lbl}). The ToAs were fitted using a model including the period $P$ and period derivative $\dot{P}$. We do not need to fit for position as it is well determined from the imaging as described in previous sections. We also do not fit for DM as this is sufficiently well determined from optimizing the signal-to-noise of the individual MeerKAT pulses.

\section*{Archival radio searches}
\label{sec:archivalradio}

\subsection*{VLITE}
\label{sec:VLITE}
The VLA Low-band Ionosphere and Transient Experiment (VLITE) \cite{2016ApJ...832...60P,2016SPIE.9906E..5BC} is a commensal instrument on the National Radio Astronomy Observatory's Karl G.\ Jansky Very Large Array (NRAO VLA) that records and correlates data across a 64 MHz bandwidth at a central frequency of 340 MHz. Since 2017-07-20, VLITE 
has been operating on up to 18 antennas during nearly all regular VLA operations, accumulating roughly 6000 hours of data per year. An automated calibration and imaging pipeline\cite{2016ApJ...832...60P}  processes all VLITE data, producing final calibrated visibility data sets and self-calibrated images. These images and associated META data are then passed through the VLITE Database Pipeline (VDP)\cite{2019ASPC..523..441P} to populate a Structured Query Language (SQL) database containing cataloged sources. 

Using VDP, we searched for all VLITE data sets that contain the position of \ulpo{} within 2.5$^\circ$ from the phase center of the VLITE observations taken when the VLA was in its A and B configurations. We note that the half-power radius of the VLITE primary beam response is $\sim$ 1.25$^\circ$, however the system is sensitive to sources well beyond this radius. We identified 124 VLITE data sets observed 
between 2017-11-09 and 2023-08-22. From these, we selected all observations with a length of at least 15 minutes, for a total of 26 observations in A configuration (angular resolution $\sim$5$^{\prime\prime}$) and 10 observations in B configuration (resolution $\sim$20$^{\prime\prime}$).
None of these 36 data sets are targeted observations of \ulpo{}, rather the source position ranges 
between a radius of 1.1$^\circ$ to 2.2$^\circ$ from the pointing center of the VLITE observation. 

To search for possible 340 MHz emission from \ulpo{} in the VLITE data, we first subtracted all known continuum sources from the self-calibrated visibilities of each observation, we then phase-shifted the data 
to the position of \ulpo{} using {\tt chgcentre}\citep{offringa-wsclean-2014}, and finally we made a time series of
dirty images of the target at 10s and 4s intervals using {\tt WSClean} \citep{offringa-wsclean-2014}. 
The primary-beam corrected noise in the 10s snapshots ranges on average from 41\,mJy\,beam$^{-1}$ (when the position of \ulpo{} is 1.1$^\circ$ away from the VLITE phase center) to 130 \,mJy\,beam$^{-1}$ (at 2.2$^\circ$). 
In the 4s snapshots, the average primary-beam corrected noise ranges between 62 \,mJy\,beam$^{-1}$ (at 1.1$^\circ$ from the phase center) and 181 \,mJy\,beam$^{-1}$ (at 2.2$^\circ$). No significant pulses from \ulpo{} were detected.

\subsection*{VLA and GMRT}
The archives of the VLA and the Giant Metrewave Radio Telescope (GMRT) were searched for data in which the position of \ulpo{} lies within the observation field-of-view. VLA P-band ($\sim$325\,MHz) observations of PSR 1937+21, 1.1$^\circ$ away, were made on 2013-09-25, 2013-11-28/2013-11-29, 2013-11-30/2013-12-01, 2013-12-01/2013-12-02 and 2014-07-07 with the array in the B and A configurations ($\sim$20$^{\prime\prime}$ and $\sim$5$^{\prime\prime}$ resolutions), respectively. The 25 Sept 2013 observation was on-source for 42 min except for short calibrator scans intermixed, while the other 2013 observations consisted of three 6-min scans separated by 1 hr, and the 2014 observation consisted of ten 6-min scans spread over 4 hr. Calibration and imaging were performed using the Astronomical Imaging Processing System \citep[AIPS;][]{Wells1985}. Amplitude and phase calibrations were both performed using 3C48, as no separate phase calibrator was observed. Imaging was performed with the AIPS task, IMAGR. The field was self-calibrated on a wide-field image using 19~facets to cover the $\sim$ 3$^\circ$ (FWHM) field-of-view. 10-s integration snapshot images were made after subtracting the deep image clean-components from the UV-data using the AIPS task, UVSUB. The RMS noise of the 10-s snapshot images at the location of \ulpo{} was typically $\sim$15\,mJy\,beam$^{-1}$ after applying a 1.5$\times$ correction for primary beam attenuation.

Two L-band ($\sim$1.4\,GHz) observations on 28-29 May 2020 (GMRT; 2.5 hr; $\sim$2$^{\prime\prime}$ resolution) and 27-Jun-2021 (VLA; 1 hr; C configuration; $\sim$10$^{\prime\prime}$ resolution) targeted SGR 1935+2154 located only $0.1^\circ$ away from \ulpo{}. A similar analysis in AIPS was followed as with the P-band observations except that wide-field imaging was not necessary due to the much smaller $\sim 0.5^\circ$ (FWHM) fields-of-view. The amplitude and phase calibrators were 3C48 and 1822-096 (GMRT) and 3C286 and J1925+2106 (VLA). The RMS noise of the 5.4-s (GMRT) and 5.0-s (VLA) snapshot images at the location of \ulpo{} were both typically $\sim$0.5\,mJy\,beam$^{-1}$ after applying a 1.1$\times$ correction for primary beam attenuation. No significant pulses from \ulpo{} were detected.

\section*{X-ray searches}
\label{sec:xray-searches}

For \textit{Chandra}, we identified 5 \textit{Chandra} observations for a total of 157.7\,ks of exposure as listed in Extended Data Table 2.
For all observations \ulpo{} was located on a front-illuminated CCD: ACIS-S2 for 21305/21306 and ACIS-S4 for the remainder.  

The observations were analyzed and combined using \texttt{CIAO} \citep[version 4.15.1, with \texttt{CALDB} 4.10.2][]{ciao}.  We first examined all observations individually for background flares by looking visually at the summed lightcurves.  No flares were identified.  We reprocessed the data to level-2 using a consistent calibration database, and reprojected the data to a common tangent point.  We then combined the reprojected observations to create an exposure-corrected image.  No source was found within $2\arcsec$ of \ulpo{}.  We also looked at the individual reprojected event files.  For each file we computed the number of events within a $2\arcsec$ radius of \ulpo{} along with the background rate determined from all of the counts on the appropriate CCD between 0.3 and 10\,keV.  A total of 6\,counts were found near the position of \ulpo{}, but the mean background rate predicts 3.8\,counts, and the chance of getting $\geq6$\,counts is 9.2\%.  Therefore we do not consider this a significant detection and place a $3\sigma$ upper limit of 10\,counts in 157.7\,ks or a count-rate limit of $6.3\times 10^{-5}\,{\rm counts\,s}^{-1}$ (0.3--10\,keV).  

Based on the observed DM of $145.8\pm3.5\,\dmunits$, we predict an absorbing column density of $N_{\rm H}\approx 4\times 10^{21}\,{\rm cm}^{-2}$ \citep{hnk13}.  We computed unabsorbed flux limits for two spectral models: a blackbody with {$kT=0.3\,$keV} (appropriate for thermal emission from a young pulsar/magnetar), and a powerlaw  with index $\Gamma=2$ (appropriate for non-thermal emission from an energetic pulsar/magnetar), following \citep{rczd+22}. This was done using \texttt{PIMMS}\footnote{\url{https://cxc.harvard.edu/toolkit/pimms.jsp}}, where we assumed a response appropriate for \textit{Chandra} cycle 22 and used the ACIS-I CCDs in place of the front-illuminated ACIS-S CCDs. With the blackbody model we infer an unabsorbed flux limit of $F_{\rm BB}<1.3\times 10^{-15}\,{\rm erg\,s}^{-1}\,{\rm cm}^{-2}$, while with the powerlaw model we infer an unabsorbed flux limit of $F_{\rm PL}<1.7\times 10^{-15}\,{\rm erg\,s}^{-1}\,{\rm cm}^{-2}$. These imply luminosity limits of $\approx 4\times 10^{30}d_{4.85}^2\,{\rm erg\,s}^{-1}$. Overall, the \textit{Chandra} observations lead to very low limits regarding the time-averaged X-ray flux. 

To search for any flaring from \ulpo{} we used extensive observations from \textit{Swift}, with  291 individual visits using the X-ray Telescope (XRT; \cite{xrt}) lasting 5--2600\,s from 2010 December through 2022 December for a total exposure time of 302.4\,ks (exposure-corrected).  We combined the individual barycentered exposures using \texttt{HEADAS} \citep{heasoft} on \texttt{SciServer} \citep{sciserver} into a single exposure-corrected dataset.  We see no sources in the summed dataset at the position of \ulpo{}: there are 33\,counts within $15\arcsec$ of \ulpo{}, but the background expectation computed using an annulus from $30\arcsec$ to $60\arcsec$ is 35.4\,counts, so we estimate a $3\sigma$ upper limit of 53.0\,counts or a count-rate limit of $1.7\times 10^{-4}\,{\rm counts\,s}^{-1}$ (0.15-10\,keV).  Using the same spectral models as above we limit the unabsorbed flux (0.3--10\,keV) to be $F_{\rm BB}<7.4\times 10^{-15}\,{\rm erg\,s}^{-1}\,{\rm cm}^{-2}$ and $F_{\rm PL}<1.1\times 10^{-14}\,{\rm erg\,s}^{-1}\,{\rm cm}^{-2}$.  These are significantly less constraining than the corresponding \textit{Chandra} limits.  

\section*{Optical and Near-Infrared searches}
\label{sec:nir}
The position of \ulpo{} was observed by the Very Large Telescope (VLT) using the near-infrared HAWKI \citep{hawki} imager.  There were a number of observations of \sgr\ that placed \ulpo{} near the edge of the field-of-view; we found the observations on 2015~April~2 to be the most useful.  These included 300\,s exposures in the $J$, $H$, and $K_s$ bands (1.2\,$\mu$m, $1.6\,\mu$m, and $2.1\,\mu$m).  However, even these exposures were not ideal, with weightmap values only 20\% of the peak at the position of \ulpo{}.  Nonetheless the collecting area of VLT makes them valuable.

We show a RGB composite of the field around \ulpo{} in Figure \ref{fig:vlt_archival}.  It is clear that the source is near the edge of the field, and is barely covered by the $H$-band image.  Coverage in $J$ and $K_s$ bands is better.  There is a source within a $1\arcsec$ radius around \ulpo{}. This source is about $0\farcs7$ away from the position of \ulpo{}, and has $J=18.4\pm0.1$\,mag, $H=17.3\pm0.1\,$mag, and $K_s=17.1\pm0.1$\,mag (Vega).  We compute the chance of finding a source randomly (drawn from the background) in the $K_s$ image with magnitude brighter than or equal to this value is 5\% (i.e. $\sim2\sigma$ association). This suggests that the association between the near-infrared source and \ulpo{} is not statistically significant.  Otherwise we infer 5$\sigma$ upper limits of $J>21.4$\,mag, $H>20.5$\,mag, and $K_s>19.8$\,mag. Aside from the deep VLT pointings, we examined images from Data Release 2 of the Pan-STARRS1 (PS1) $3\pi$ survey \citep{ps1}.  There is a source in the ``stack" catalog that corresponds to the near-IR source identified above.  This source is cataloged as PSO J293.7711+21.8119.  The source is not detected in the $g$, $r$, or $i$-bands, and has only detections in $z$ ($22.0\pm0.3$\,mag) and $y$ ($20.4\pm0.1$).  For the other bands and for the rest of the error region we adopt the standard PS1 stack upper limits $g>23.3$, $r>23.2$, $i>23.1$, $z>22.3$, and $y>21.3$.

\section*{Search for persistent radio emission}

To ascertain if there is an un-pulsed radio component which might be attributed to a wind nebula, or perhaps indicate emission of a non-neutron star origin, we imaged the ASKAP visibility data.  In a stacked ASKAP deep image at 887.5~MHz, if we subtract the mean background emission there is no detection above 3$\sigma$ with an rms of 25 $\mu$Jy/beam. However, the location of the source at the edge of the supernova remnant in Extended Data Figure 2 makes it is difficult to determine exactly what background emission to subtract. Therefore we are unable to confirm the presence of potential (possibly faint) persistent continuum radio emission in either the MeerKAT or ASKAP data.

\section*{Faraday conversion}

We tested whether the large circular polarisation fraction seen in the MeerKAT beamformed data could be due to Faraday conversion using a simple phenomenological model \citep{ljl+23}, where the polarisation vector is modelled as a series of frequency-dependent rotations on the Poincar{\'e} sphere. However, we failed to recover any significant frequency-dependence to the circular polarisation. This indicates it is either intrinsic to the emission mechanism or arises from a more complex propagation effect such as the partially coherent addition of linearly polarised modes \citep{okj+23}.

\section*{Model constraints}

\subsection*{White Dwarf}
\label{sec:WDconstraints}

We examine the potential for the optical/near-IR data described above to constrain white dwarf scenarios for the source.  We used the synthetic photometry\footnote{See \url{https://www.astro.umontreal.ca/~bergeron/CoolingModels/}.} of \citep{2011A&A...531L..19T,2018ApJ...867..161B,2020ApJ...901...93B} together with the 3D extinction model of \citep{2019ApJ...887...93G} to compute distance constraints as a function of effective temperature for hydrogen-atmosphere (DA) white dwarfs with masses $0.6\,M_\odot$ and $1.0\,M_\odot$, representing standard and massive white dwarfs, respectively (using helium atmosphere DB white dwarfs does not change the conclusions). We did two analyses, one where we modeled the potential near-IR counterpart and one where we treated the source as non-detections. Note that there are large degeneracies involved: extinction and effective temperature are highly degenerate, and other quantities such as mass (and hence radius) degenerate with distance. When considering the potential near-IR counterpart as correct, and  given the sparse data that would all be on the Rayleigh-Jeans tail, we unsuprisingly found a plausible fit to the VLT data with $T_{\rm eff}\approx 15000\,$K and distance $\approx 6\,$kpc (implying reddening $E(B-V)\approx 2.8$). However, the implied radius is $\approx 0.8\,R_\odot$, leading us to conclude that this source cannot be expected by standard white dwarf models.

Considering only upper limits (so assuming that the source PSO~J293.7711+21.8119 is \textit{not} associated with \ulpo{}) we find that a white dwarf with $T_{\rm eff}<30000\,$K could be present at distances $>1\,$kpc.
Despite their greater sensitivity, we found the VLT data generally less constraining than the PS1 data given the range of effective temperatures considered. Given the implied average DM distance of 4.85~kpc from the NE2001 and YMW16 models, we do not consider the limits described here especially constraining. 
Under the framework of coherent radio emission from pair production, the compactness of the source can be related to the period and 
magnetic field so that we can estimate the minimum radius of the source \cite{bwh+23} using,

\begin{equation}
    R \gtrsim 4 \times 10^9 \Big(\frac{Q_{c}}{10}\Big)^{4/17}\,\Big(\frac{P}{1000~\rm{s}}\Big)^{13/17}\,\Big(\frac{B}{10^9~\rm{G}}\Big)^{-8/17}\,\rm{cm},
\end{equation}

\noindent where $P$ is the period in seconds, $B$ is the magnetic field in Gauss and $Q_{c} = \rho_{c}/R$ is the dimensionless characteristic field curvature radius in which the curvature radius near the polar cap is assumed to be $\rho_{c} \sim 10R$. The conventional emission model for any compact object to emit dipole radiation assumes the existence of a vacuum gap above the polar cap. In order to sustain pair production, the potential difference across this gap must be sufficiently large and this is no longer possible beyond the classic death line \citep{CR93}. As a result, pair production and consequently, radio emission ceases. The relation above therefore encodes radio deathline physics due to requirements on pair-cascade production and provides a lower-limit on the compactness of an object to sustain this emission. A choice of $Q_{c} \gg 10$ is commensurate with the expected small polar cap size for a $P\sim1000$~s rotator. Assuming $10 \leq Q_{c} \leq 10^5$ and $B = 10^9$~G, we can rule out an isolated magnetic white dwarf origin for the observed emission as the estimated lower-limit on the radius of 0.14$R_\odot$, even in the case of $Q_{c} = 10$, is much too large for a white dwarf (Extended Data Figure 5).

\subsection*{Neutron Star}
\label{sec:NSconstraints}
 
It has been proposed that bright coherent radio bursts can be produced by highly magnetized neutron stars that have attained long rotation periods (few 10s to a few 1,000s of seconds), called ultra-long-period magnetars (ULPMs). 
Typically, magnetars have quiescent  X-ray luminosities anywhere between $10^{31}$ and $10^{36}\,{\rm erg\,s}^{-1}$ \citep{2014ApJS..212....6O, 2019MNRAS.485.4274H} regardless of radio emission (typically they are brighter in X-rays following outbursts that lead to radio emission), and so our deep X-ray limits from searching archival X-ray data challenge the magnetar interpretation.  However, there are sub-classes of magnetars with considerably weaker X-ray emission, $<10^{30}\,{\rm erg\,s}^{-1}$, such as the ``low-field" magnetars whose \citep{2013IJMPD..2230024T} spin-down inferred fields are $\sim 10^{13}\,$G, but where local X-ray absorption features suggest much higher localized fields \citep{2013Natur.500..312T,2016MNRAS.456.4145R}.  If indeed \ulpo{} and similar sources are part of another emerging sub-class of magnetars, the quiescent X-ray luminosities (which are attributed to the decay of large-scale dipole magnetic fields) may be lower. If that is the case, it would also explain the location of \ulpo{} as magnetars are typically expected to be young objects that lie in the Galactic Plane \citep{kb17}. Combining all these observational aspects, \ulpo{}, is likely part of an older population of magnetars with long spin-periods, low X-ray luminosities but magnetized enough to be able to produce coherent radio emission. It is important that we probe this hitherto unexplored region of the neutron star parameter space in order to get a complete picture of the evolution of neutron stars, and this may an important source to do so.


\backmatter

\section*{Declarations}

\begin{itemize}

\item \textbf{Funding}
M.C. acknowledges support of an Australian Research Council Discovery Early Career Research Award (project number DE220100819) funded by the Australian Government. Parts of this research were conducted by the Australian Research Council Centre of Excellence for Gravitational Wave Discovery (OzGrav), project number CE170100004. R.M.S. and N.H.W. acknowledge support through Australian Research Council Future Fellowships FT190100155, and FT190100231, respectively. T.E.C. and S.G. acknowledge that basic research in Radio Astronomy at the U.S.\ Naval Research Laboratory is supported by 6.1 Base Funding. KMR acknowledges support from the Vici research programme ``ARGO'' with project number 639.043.815, financed by the Dutch Research Council (NWO). 

\item \textbf{Data availability} 
The data that support the findings of this study are available at Zenodo: \url{https://doi.org/10.5281/zenodo.10989868}

\item \textbf{Code availability} The timing was performed using \textsc{tempo2} \citep{tempo2}. Specific Python scripts used in the data analysis are available on request from M.C.

\item \textbf{Acknowledgements} M.C. would like to thank Elaine Sadler, Ron Ekers, Dan Huber, Lucy Oswald and Stefan Oslowski for valuable discussions. This manuscript makes use of data from MeerKAT (Project ID: DDT-20210125-MC-01) and Parkes (Project ID: PX079). M.C. would like to thank SARAO for the approval of the MeerKAT DDT request, the science operations, CAM/CBF and operator teams for their time and effort invested in the observations, and the ATNF for scheduling observations with the Parkes radio telescope.
The MeerKAT telescope is operated by the South African Radio Astronomy Observatory, which is a facility of the National Research Foundation, an agency of the Department of Science and Innovation (DSI). 
TRAPUM observations used the FBFUSE and APSUSE computing clusters for data acquisition, storage, and analysis. These clusters were funded and installed by the Max-Planck-Institut f{\"u}r Radioastronomie and the Max-Planck-Gesellschaft. 
This scientific work uses data obtained from telescopes within the Australia Telescope National Facility (\url{https://ror.org/05qajvd42}), which is funded by the Australian Government for operation as a National Facility managed by CSIRO.
The Parkes radio telescope (Murriyang) is part of the Australia Telescope National Facility (\url{https://ror.org/05qajvd42}), which is funded by the Australian Government for operation as a National Facility managed by CSIRO. 
We acknowledge the Wiradjuri people as the Traditional Owners of the observatory site.  Inyarrimanha Ilgari Bundara / the Murchison Radio-astronomy Observatory is the site of the CSIRO ASKAP radio telescope. We acknowledge the Wajarri Yamaji as the Traditional Owners and native title holders of the observatory site. Operation of ASKAP is funded by the Australian Government with support from the National Collaborative Research Infrastructure Strategy. ASKAP uses the resources of the Pawsey Supercomputing Research Centre. Establishment of Inyarrimanha Ilgari Bundara, the CSIRO Murchison Radio-astronomy Observatory, ASKAP and the Pawsey Supercomputing Research Centre are initiatives of the Australian Government, with support from the Government of Western Australia and the Science and Industry Endowment Fund.
This paper includes archived data obtained through the CSIRO ASKAP Science Data Archive, CASDA (\url{http://data.csiro.au}). The National Radio Astronomy Observatory is a facility of the National Science Foundation operated under cooperative agreement by Associated Universities, Inc. Construction and installation of VLITE was supported by the NRL Sustainment Restoration and Maintenance fund.

\item \textbf{Author contributions}
M.C. drafted the manuscript with suggestions from all co-authors and is the PI of the MeerKAT data. M.C. reduced and analysed the MeerKAT TUSE/PTUSE data and undertook the timing analyses along with R.M.S. E.L. calibrated, imaged and performed astrometry on the ASKAP and MeerKAT data. D.L.K, T.M., L.F, K.M.R, C.M.L.F, J.W.T, M.K., J.P., and B.W.S. contributed to discussions about the nature of the source. Y.P.M analysed the MeerKAT APSUSE data and performed the beam subtraction. S.G and T.E.C. performed the VLITE archive search, imaging and analyses. S.D.H. performed the VLA and GMRT archive search, imaging and analyses. M.E.L. performed the Faraday conversion analysis and is the PI of the Parkes PX079 project. V.R. performed the Keck observations and calibration of the optical data. E.D.B. built and designed the beamformer used by MeerTRAP. S.B. scheduled the MeerKAT observations. C.M.L.F. interpreted the optical spectrum along with M.C. B.W.S is PI of MeerTRAP.

\item \textbf{Conflict of interest/Competing interests} 
The authors declare no competing interests.

\end{itemize}

\clearpage




\begin{table}
\footnotesize
\centering
\caption{Measured and derived radio quantities for \ulpo{} from the ASKAP and MeerKAT observing campaigns. Uncertainties are 1-$\sigma$ errors on the last significant quoted digit. The best fit coordinates in the table are from the MeerKAT localisation and the quoted beamformed time resolution corresponds to the best time resolution of all the backend instruments used (see Methods for more details).}
\begin{tabular}{lll} 
\hline \\
\textbf{Parameter}                               & \textbf{ASKAP}                 & \textbf{MeerKAT} \\ \\ \hline \\
Centre frequency                        & 887.5~MHz             & 1284~MHz \\
Bandwidth                               & 288~MHz               & 856~MHz \\
Imaging time resolution                 & 10~seconds            & 2~seconds \\
Beamformed time resolution              & --                    & $38.28\,\mu$s \\
Typical widths, $W$                     & 10 - 50~seconds       & $\sim370$~ms \\ 
Linear Polarisation fraction, $L/I$     & $>90\%$               & $\sim40\%$ \\
Circular Polarisation fraction, $V/I$   & $<3\%$                & $>70\%$ \\
Inband spectral index, $\alpha$         & $+0.4\pm0.3$          & $-1.2\pm0.1$ \\
Rotation Measure, RM                    & $+159.3\pm0.3$~\rmunits        & $+159.1\pm0.3$~\rmunits \\
Peak flux density of brightest pulse, $S_{\nu}$            & 234.7~mJy               & 9~mJy \\
Radio luminosity, $L_{\nu}$             & $1.8\times10^{30}\rm{erg\,s^{-1}}$ & $2.1\times10^{29}\rm{erg\,s^{-1}}$ \\
Imaging timescale                       & 10~seconds            & 2~seconds \\
Epoch                                   & 2022-10 to 2023-02    & 2023-02 to 2023-08 \\
Dispersion measure, DM                  & --                    & $145.8\pm3.5$~\dmunits \\ \hline \\ 
Right Ascension (J2000)                 & \multicolumn{2}{c}{$19^h35^m05.175^s \pm 0.3''$}  \\
Declination (J2000)                     & \multicolumn{2}{c}{$+21^d48'41.504'' \pm 0.6''$} \\ 
Period                                  & \multicolumn{2}{c}{$3225.309\pm0.002$~seconds} \\     
Period derivative                       & \multicolumn{2}{c}{$\leq(1.2\pm1.5)\times10^{-10}$~s~s$^{-1}$}\\
Distance (\textsc{ymw16}), $d_{1}$      & \multicolumn{2}{c}{$4.3$~kpc}\\
Distance (\textsc{ne2001}), $d_{2}$     & \multicolumn{2}{c} {$5.4$~kpc}\\
Neutron star surface dipole magnetic field strength & \multicolumn{2}{c} {$\leq \mbox{a few}\times 10^{16}$\,G}\\
White dwarf surface dipole magnetic field strength  & \multicolumn{2}{c} {$\leq \mbox{a few}\times 10^{10}$\,G}\\
Spin-down luminosity (white dwarf), $\dot{E}$ & \multicolumn{2}{c} {$\leq1.4\times10^{31}$~erg~$\mathrm{s}^{-1}$} \\
Spin-down luminosity (neutron star), $\dot{E}$  & \multicolumn{2}{c} {$\leq1.4\times10^{26}$~erg~$\mathrm{s}^{-1}$} \\

\hline
\end{tabular}
\label{tab:params}
\end{table}


\setcounter{table}{0}
\captionsetup[table]{name={\bf Extended Data Table}}

\begin{landscape}
\begin{table*}
\footnotesize
\centering
\caption{ASKAP and MeerKAT observations of \ulpo{}. The first three rows labelled Epoch 0 are archival observations targeting \sgr{}, while the rest are follow-up observations targeting \ulpo{}. See the text for details.}
\begin{tabular}{llllllll} 
\toprule
Epoch       & SBID               & Start Time             & Duration       & Frequency & Observation     & Peak flux density      & Note  \\
            &                    & UT, J2000              & UT, J2000      & MHZ              &                 & mJy                    &        \\
\midrule
0           & 20200507-0039      & 2020-05-08 05:16:47    & 01:16:49       & 1284             & MeerKAT         & --                     & no detection. \\  
0           & 20200510-0034      & 2020-05-11 03:01:01    & 03:17:30       & 1284             & MeerKAT         & --                     & no detection. \\ 
0           & 20200514-0008      & 2020-05-15 00:40:59    & 03:19:20       & 1284             & MeerKAT         & --                     & no detection. \\ 
1           & 44780              & 2022-10-12 07:00:00    & 06:02:35       & 887.5            & AS113\_66       & 6.9                    & one 7$\sigma$. \\
2           & 44857              & 2022-10-15 07:15:02    & 06:02:36       & 887.5            & AS113\_67       & 118.8, 93.7, 18.4,     & Discovery - four $>6\sigma$; one $>4\sigma$. \\
            &                    &                        &                &                  &                 & 17.9, 13.8             & \\
3           & 44918              & 2022-10-17 06:31:08    & 03:01:27       & 887.5            & AS113\_68       & 10.8                   & one $>10\sigma$. \\
4           & 45060              & 2022-10-22 06:31:34    & 06:02:05       & 887.5            & AS113\_69       & 17.7, 10.2             & one $>10\sigma$ sigma; one $>4\sigma$. \\
5           & 45086              & 2022-10-23 06:13:17    & 06:02:19       & 887.5            & AS113\_70       & 234.7, 209.3,          & five $>6\sigma$. \\
            &                    &                        &                &                  &                 & 170.7, 146.6, 112.5    & \\
6           & 45416              & 2022-11-05 05:01:15    & 06:02:00       & 887.5            & AS113\_71       & 4.0                    & one $>4\sigma$. \\
7           & 46350              & 2022-12-13 02:27:01    & 05:23:05       & 887.5            & CRACOTest\_A    & --                     & no detection. \\ 
8           & 46419              & 2022-12-14 02:16:02    & 06:02:08       & 887.5            & CRACOTest\_A    & --                     & no detection.  \\
9           & 46492              & 2022-12-15 04:15:00    & 05:01:49       & 887.5            & CRACOTest\_B    & --                     & no detection.  \\
10          & 46554              & 2022-12-16 02:07:02    & 05:02:10       & 887.5            & CRACOTest\_B    & --                     & no detection.  \\
11          & 20230203-0012      & 2023-02-03 09:55:15    & 01:00:12       & 1284             & MeerKAT         & 9.0, 2.3               & DDT - one pulse and weak sub-pulse. \\
12          & 47635              & 2023-02-04 04:20:38    & 01:02:24       & 887.5            & ULP2            & --                     & no detection.  \\
13          & 48611              & 2023-02-25 21:33:58    & 06:03:32       & 887.5            & CRACO\_ULP2     & --                     & no detection.  \\
14          & 20230302-0029      & 2023-03-04 06:48:46    & 00:59:56       & 1284             & MeerKAT         & --                     & DDT - no detection.  \\
15          & 20230409-0012      & 2023-04-10 03:06:57    & 00:59:49       & 1284             & MeerKAT         & --                     & DDT - no detection.  \\
16          & 20230508-0002      & 2023-05-08 01:41:45    & 01:00:10       & 1284             & MeerKAT         & 2.9, 1.1               & DDT - one pulse and weak sub-pulse. \\
17          & 20230821-0011      & 2023-08-21 16:46:25    & 01:00:12       & 1284             & MeerKAT         & --                     & DDT - no detection.  \\
\bottomrule
\end{tabular}
\label{tab:observations}
\end{table*}
\end{landscape}


\begin{table}[htbp]
    \footnotesize
    \centering
    \caption{Archival \text{Chandra} observations}
    \begin{tabular}{llllll} 
        \toprule
        ObsID & Date & DOI \\
        \midrule
        21305 & 2019-11-24 & \doi{10.25574/21305}1 \\
        21306 & 2019-12-03 & \doi{10.25574/21306} \\
        22431 & 2020-04-30  & \doi{10.25574/22431} \\
        22432 & 2020-05-02 & \doi{10.25574/22432} \\
        23251 & 2020-05-18 & \doi{10.25574/23251} \\
        \bottomrule
    \end{tabular}
    \label{tab:chandraObs}
\end{table}


\clearpage



\begin{figure}
\centering
  \includegraphics[width=3.5 in]{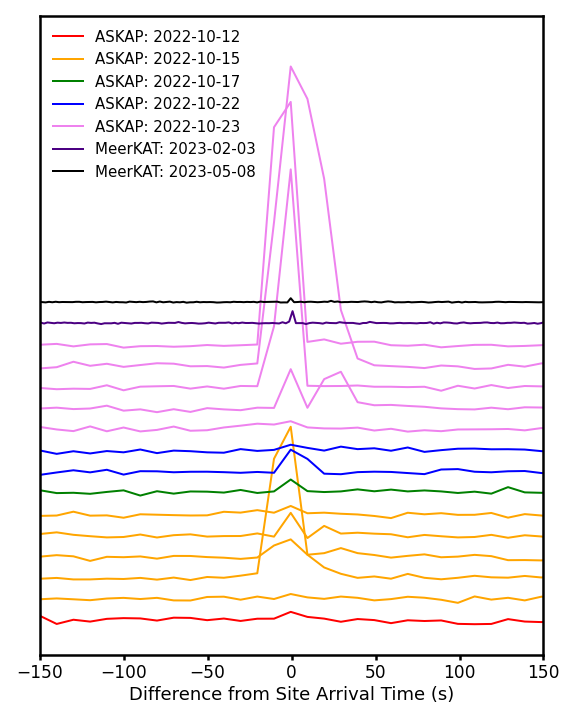}
    \caption{Light curves of the ASKAP and MeerKAT detections where the y-axis is the pulse number. The peak flux densities of these detections are reported in Extended Data Table 1. The different colours represent the dates of the observations. Pulse detections within an observation represent consecutive rotations of the source. Though different in terms of radio properties, the MeerKAT detections appear to arrive in phase with the ASKAP detections.}
\label{fig:lightcurves}
\end{figure}

\clearpage


\begin{figure}
    \includegraphics[width=2.5 in]{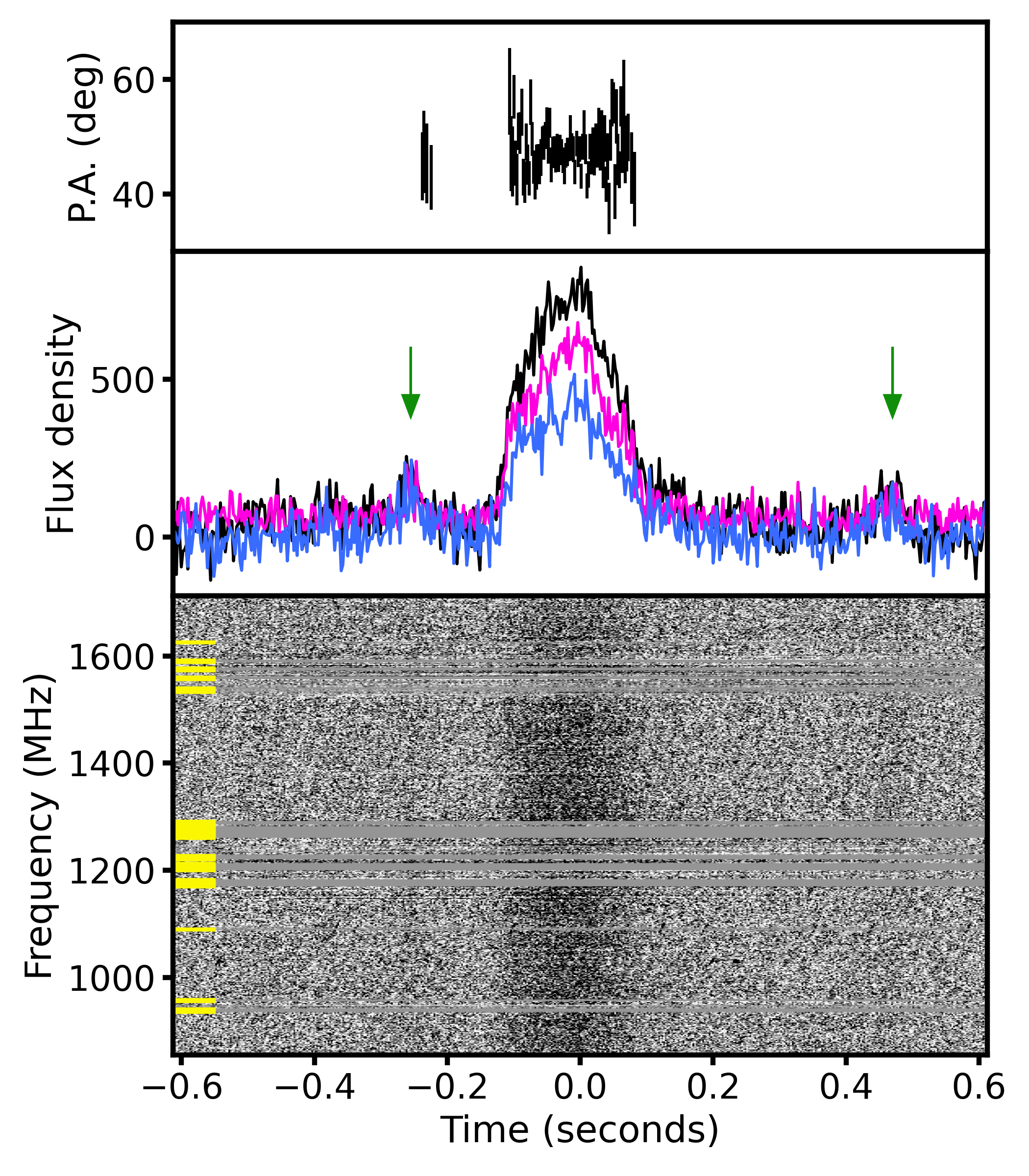}
    \includegraphics[width=2.5 in]{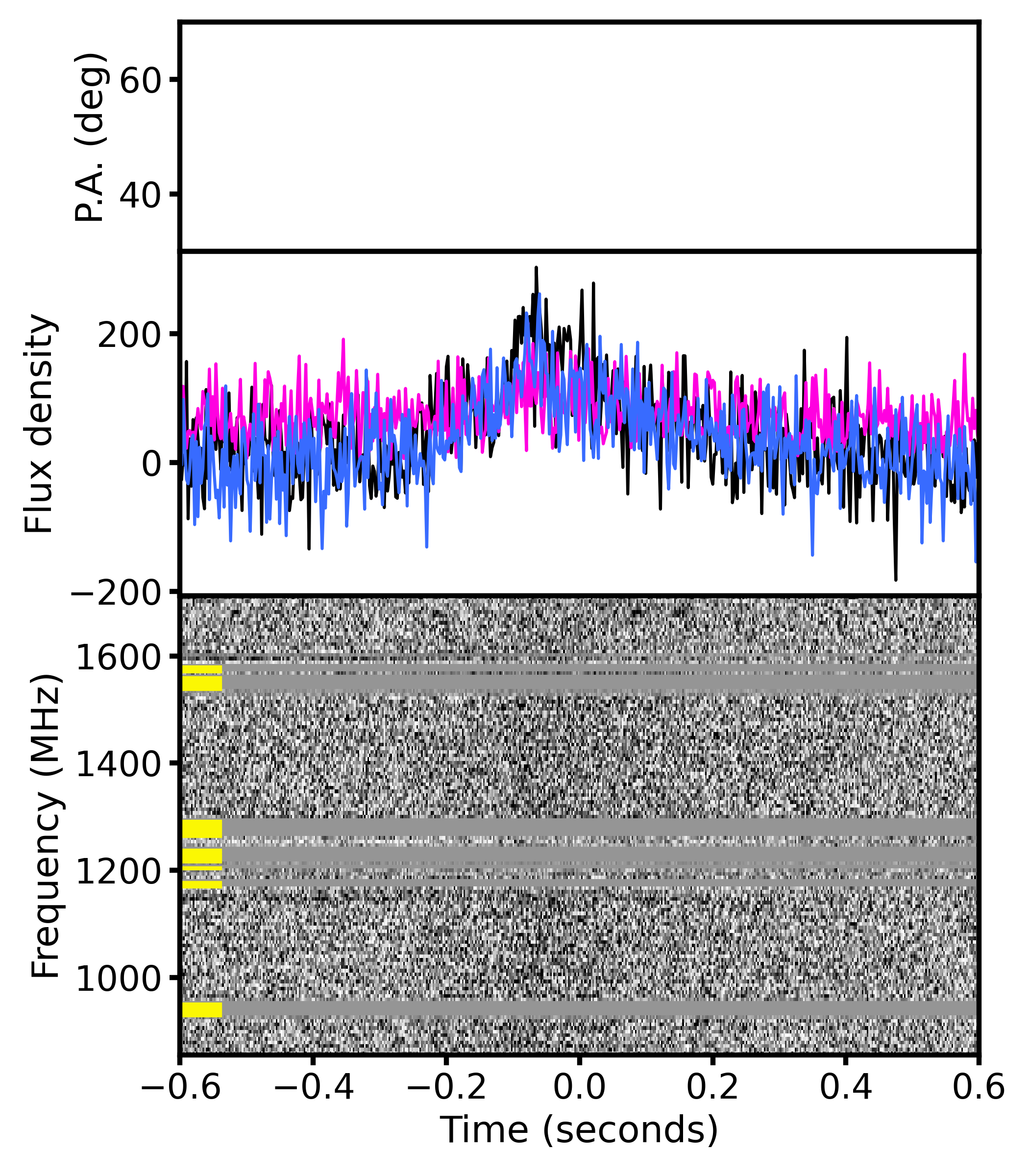}
    \caption{The dynamic spectra and polarisation pulse profiles of \ulpo{} from the MeerKAT beamformed data. The left panel shows a bright detection on 2023-02-03 and the right panel shows a weaker detection on 2023-05-08. The data have a time resolution of 2.4~ms and are coherently de-dispersed to a DM of 145.8\,$\dmunits$ and corrected for an RM of $+159.3\,\rmunits$. The top panel shows the polarisation position angle (for values of linear polarisation greater than 3 times the off-pulse noise), which is observed to be flat across the main pulse profile in the left panel. The insufficient signal-to-noise ratio during the detection on 2023-05-08, prevented robust measurements of polarisation position angles. The middle panel shows the Stokes parameter pulse profiles for \ulpo{} at 1284~MHz where black represents the total intensity, magenta represents linear polarisation and blue represents circular polarisation. The flux density is in arbitrary units as the data are not flux calibrated. The arrows indicate the positions of the pre- and post-cursor bursts for the detection on 2023-02-03 (left panel). The bottom panel shows the dynamic spectra where the backwards sweeping striations across the observing band in the left panel correspond to $\sim50$~Hz radio frequency interference.}
    \label{fig:pulse-profiles}
\end{figure}

\clearpage


\begin{figure}
\centering
  \includegraphics[width=4.5in]{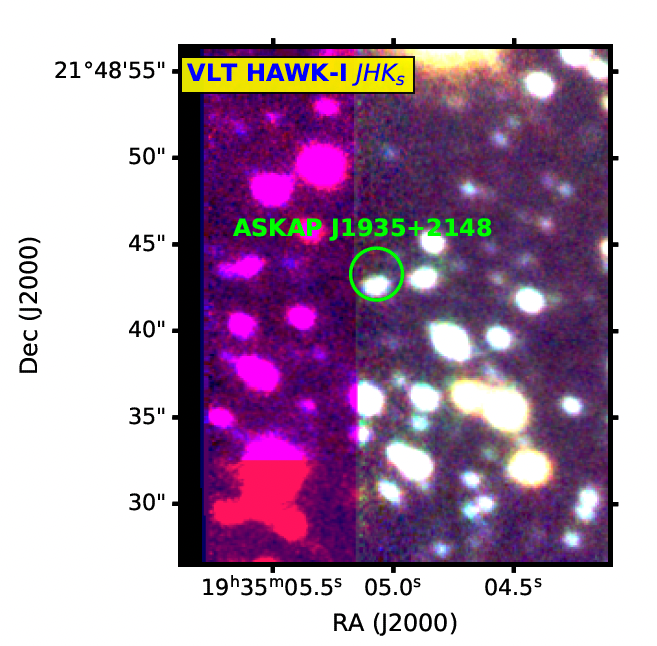}
    \caption{Three-color near-infrared image of the field around \ulpo{}, from VLT/HAWK-I $JHK_s$ imaging on 2015 April~2. The image cutout is $20\arcsec \times 30\arcsec$. The uncertainty in the position of \ulpo{}\ is shown by a green circle of radius $1\arcsec$.}
\label{fig:vlt_archival}
\end{figure}

\clearpage




\setcounter{figure}{0}
\captionsetup[figure]{name={\bf Extended Data Figure}}

\begin{figure}
\centering
 \includegraphics[width=4.0 in]{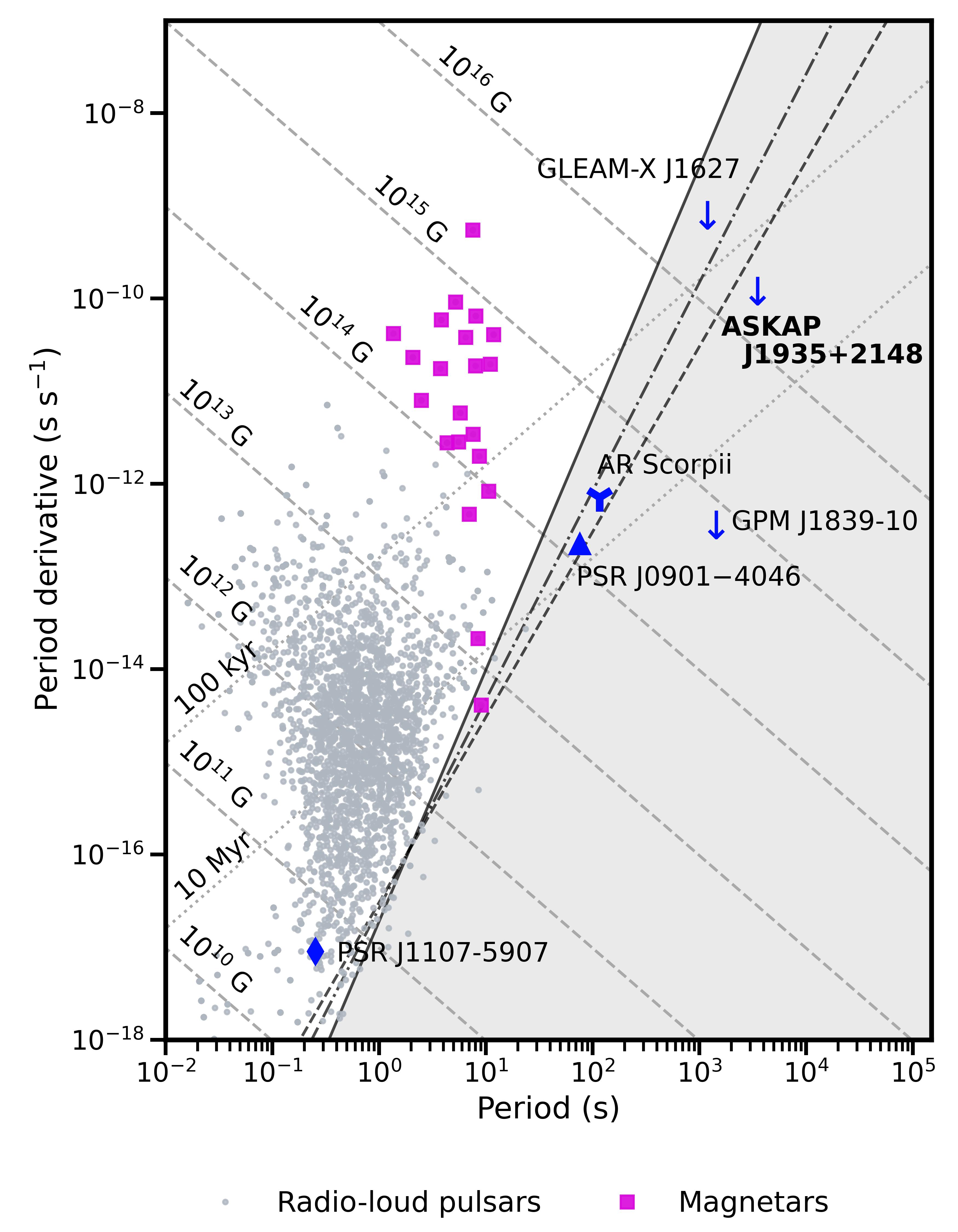}
    \caption{$P-\dot{P}$ diagram showing the spin-period against the period derivative for neutron stars as reported in the ATNF pulsar catalog, and published ultra-long period transients. Lines of constant age and magnetic field for neutron stars are shown as dotted and dashed lines respectively. The lower right region of the figure bounded by the various death lines represents the `death valley' where sources below these lines are not expected to emit in the radio. The solid death line represents Equation 9 in \citep{CR93}. In dot-dashed and dashed are the death lines modeled on curvature radiation from the vacuum gap and SCLF models as shown by Equations 4 and 9 respectively in \citep{ZHM2000}.}
\label{fig:ppdot}
\end{figure}


\begin{figure}
\centering
 \includegraphics[width=5 in]{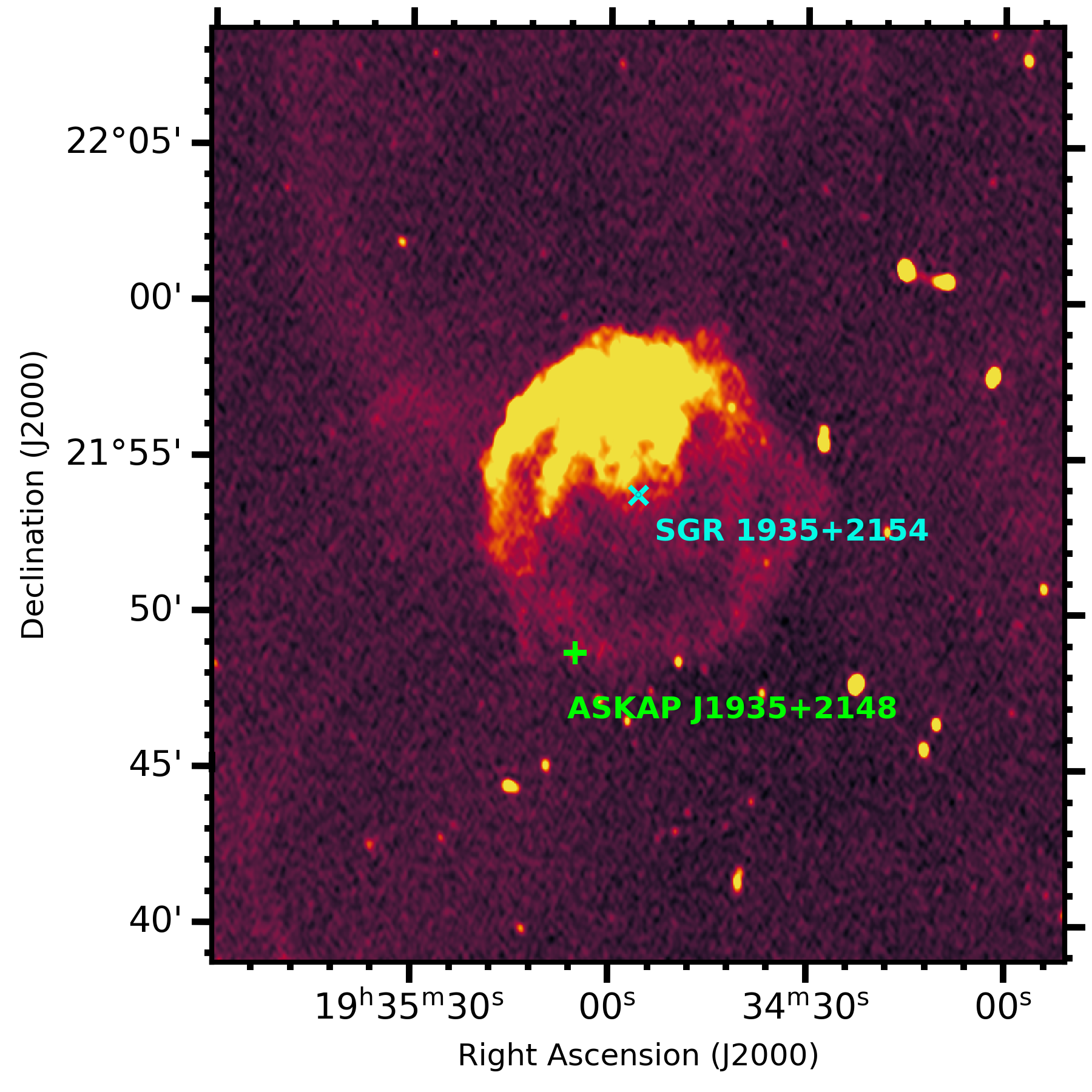}
    \caption{ASKAP deep image centered on \sgr{} spanning 6~hours with a median rms of 42~$\mu$Jy/beam. The position of \ulpo{} places it 5\arcmin6 from \sgr{}, and the DM = $145.8\pm3.5$ indicates that it is in the foreground.}
\label{fig:askap-deepimage}
\end{figure}

%
\begin{figure}
\centering
  \includegraphics[width=5 in]{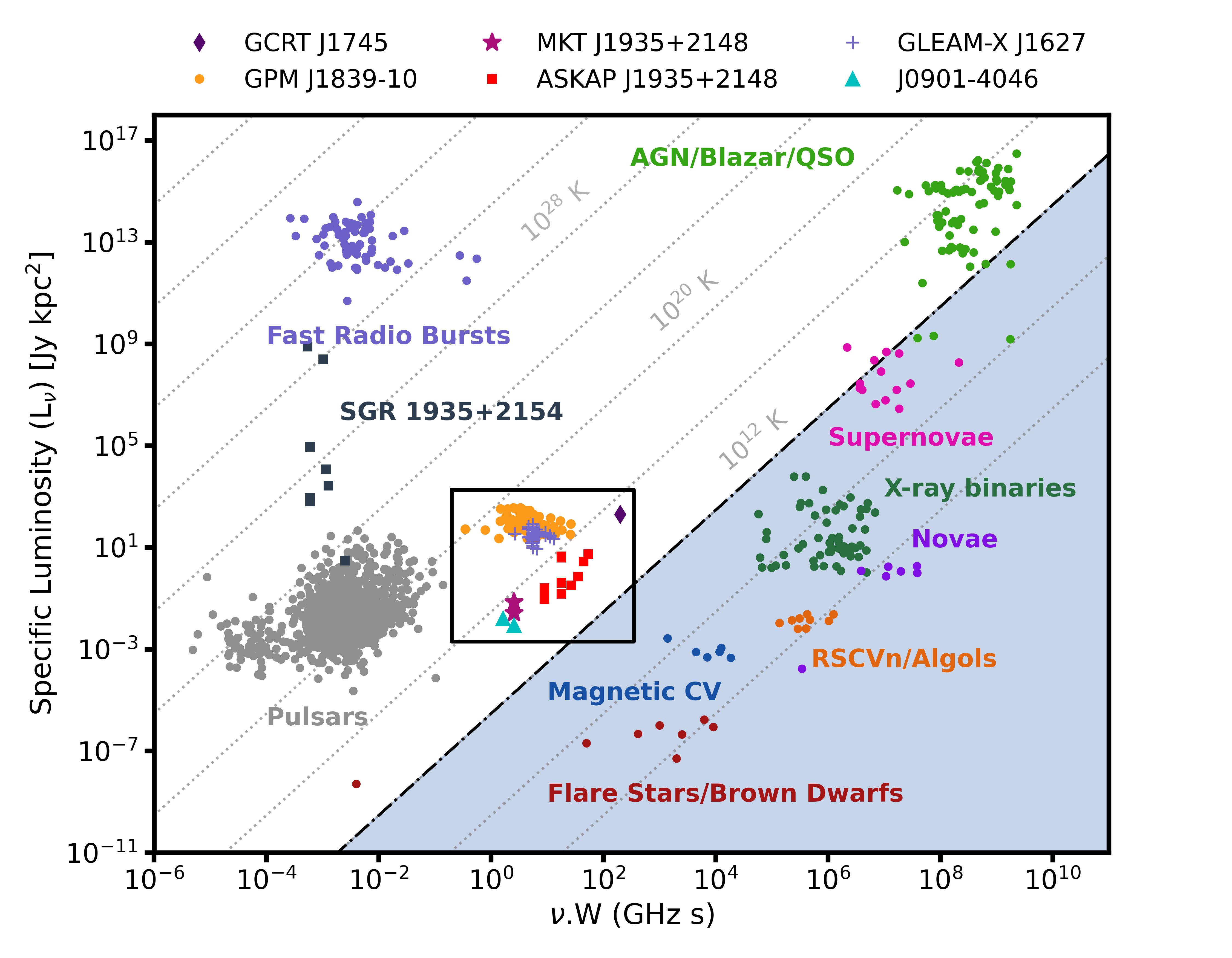}
    \caption{The luminosity of different types of transients as a function of their width ($W$) and frequency ($\nu$). Diagonal lines represent constant brightness temperatures. The brightness temperature of $10^{12}$~K separates coherent emitters from the incoherent ones, with the shaded region (lower right triangle) housing the incoherent emitters. The long period sources shown in the legend appear to cluster together as indicated by the box, which is merely to highlight the sources and does not have a physical significance.}
\label{fig:phasespace}
\end{figure}

%

\begin{figure}
\centering
 \includegraphics[width=5 in]{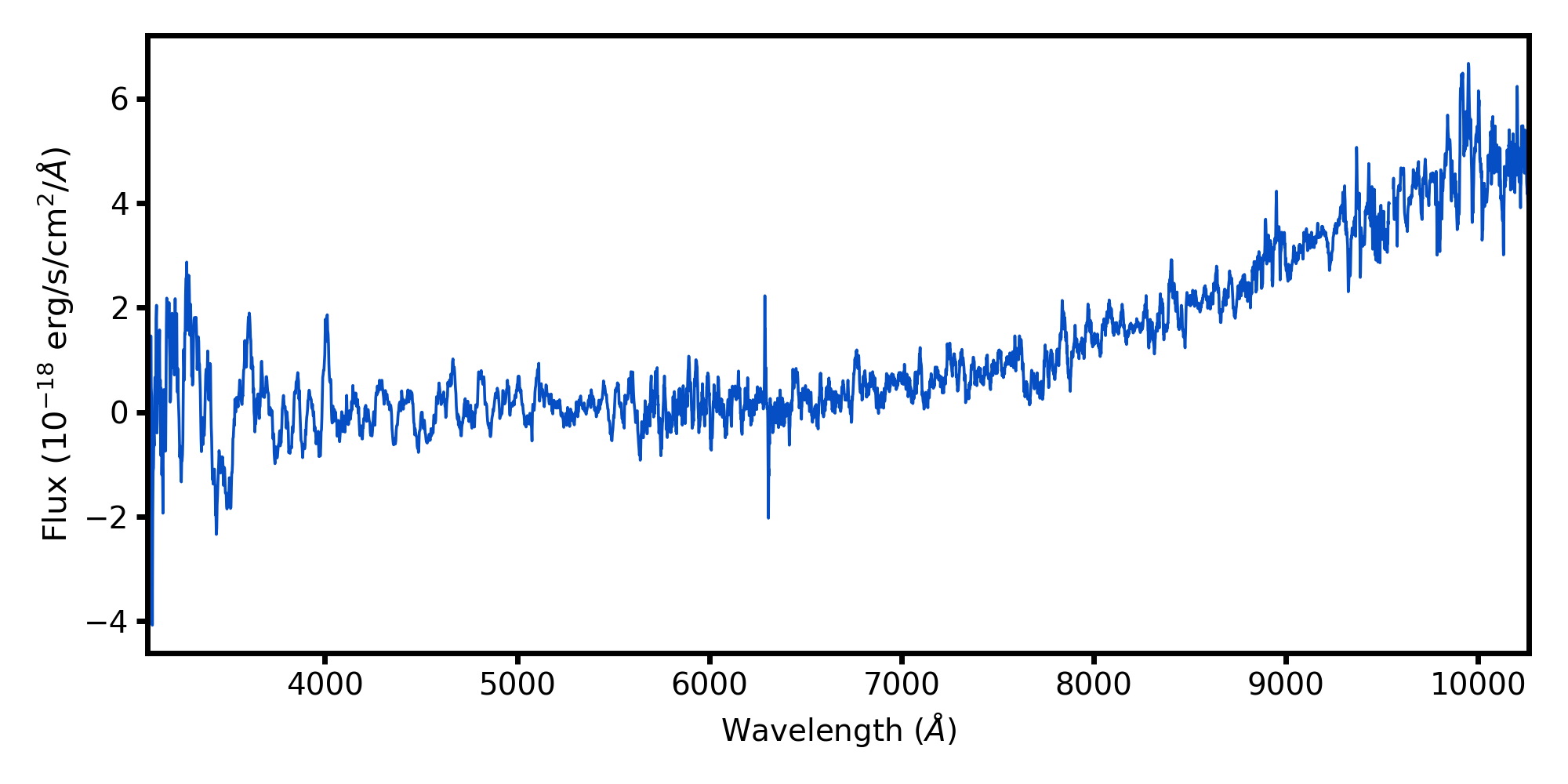}
    \caption{Spectrum of PSO\,J293.7711+21.8119 using the LRIS instrument at the Keck telescope. The data shows a featureless red continuum spectrum expected of L/T-dwarf stars.}
\label{fig:lrisspectrum}
\end{figure}


\begin{figure}
\centering
  \includegraphics[width=4 in]{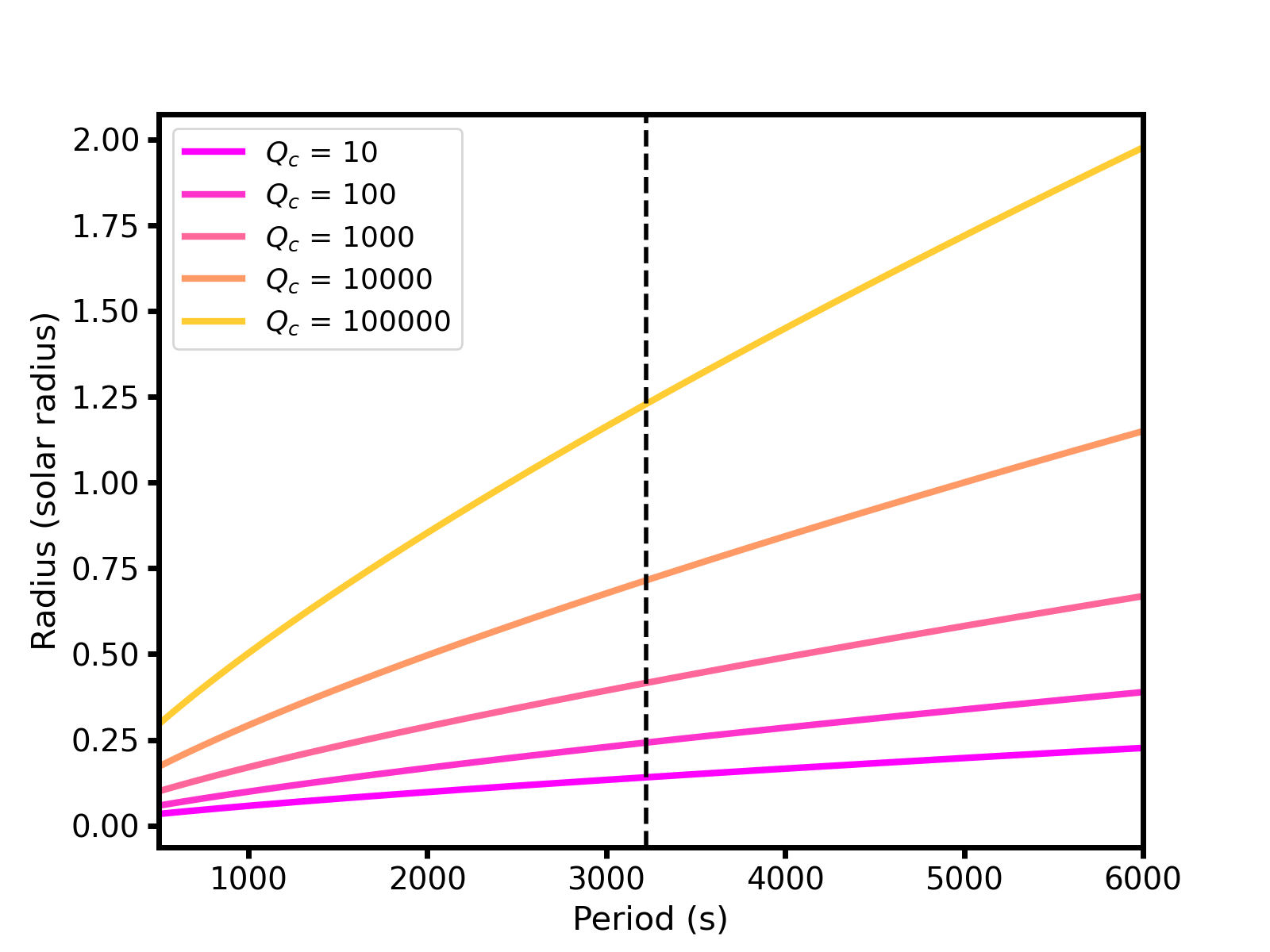}
    \caption{Constraints on the radius of a source, in units of solar radii, for various assumed rotational periods. $Q_c$ is the ratio of the field curvature radius to the stellar radius with the value inversely proportional to the size of the star. $Q_c$ is typically assumed to be 10 for WDs but is larger in reality. The vertical line denotes the period of \ulpo{}. Even in the most conservative case of $Q_c = 10$, we are able to rule out a white dwarf origin scenario. More details in Methods.}
\label{fig:RvsP}
\end{figure}


\clearpage


\end{document}